\documentclass[preprint,12pt]{elsarticle}%
\usepackage{amssymb}
\usepackage{amsfonts}
\usepackage{amsmath}
\usepackage{graphicx}%
\providecommand{\U}[1]{\protect\rule{.1in}{.1in}}
\journal{journal}

\begin{document}
\bigskip\bigskip%
\begin{frontmatter}%


%

\title
{Some further validations and comparison of the Bearing Area Model (BAM) for adhesion of rough surfaces}%

%

\author{M. Ciavarella}%
%

\address
{Politecnico di BARI. Center of Excellence in Computational Mechanics. Viale Gentile 182, 70126 Bari. Mciava@poliba.it}%
%

\begin{abstract}%

In the present short note, we attempt further validations and comparisons of a
recent simple model for the estimate for adhesion between elastic (hard) rough
solids with Gaussian multiple scales of roughness, BAM (Bearing Area Model)
belonging to a DMT class of models. In one case, we use the GJP (Generalized
Johnson\ Parameter) model, which is an empirical fit validated on the same
(and so far most extensive) set of data on which BAM was validated, namely
that of Pastewka and Robbins. In the second case, we compare with another
approximate DMT theory, that of Persson and Scaraggi, which turns out
extremely close to the BAM model, despite much more complex: GJP however can
lead to significant discrepancies.%

\end{abstract}%
%

%

\end{frontmatter}%



\section{\bigskip Introduction}

The science of adhesion starts with adhesion of "hard" solids, neglecting any
elasticity, Bradley found in 1932 (Bradley, 1932) the first pull-off force
between a spherical particle and a flat surface. The second solution only came
some 40 years later, with JKR (from Johnson et al., 1971), and that solution
differed only by a small prefactor from the Bradley solution, indicating in
that case (of quadratic surfaces), almost no influence of the elastic
behaviour, except that the elastic behaviour manifested itself in
instabilities for pull-in and pull-off which provoke hysteresis (the
hysteresis of the JKR solution was even very recently further investigated by
Ciavarella et al (2017)). Since then, many more aspects have been studied and
there has been an explosion of interest in adhesion, motivated also by the
quest for small scale engineering, bioengineering and bio-inspired
engineering. We know today that for \textquotedblleft soft\textquotedblright%
\ bodies, adhesion shows instabilities like in the simple case of a single
sinusoid in the so called JKR regime (Johnson 1995) which leads to a very
strong hysteretic behavior, with wear adhesion resulting in some cases and
strong adhesion in others, depending on a single parameter, $\alpha_{1}$,
defined in terms of surface energy $w$
\begin{equation}
\alpha_{1}=\sqrt{\frac{2}{\pi^{2}}\frac{w\lambda}{E^{\ast}h^{2}}%
}\label{alfa-KLJ}%
\end{equation}
for a single sinusoid (this is why we use the subscript "1") of wavelength and
amplitude $\lambda,h$, and where $\alpha$ represents the square of the ratio
of the surface energy in one wavelength to the elastic strain energy when the
wave is flattened. Here, $E^{\ast}$ is elastic modulus in plane strain, where
$1/E^{\ast}=1/E_{1}^{\ast}+1/E_{2}^{\ast}$ and $E_{i}^{\ast}=E_{i}/\left(
1-\nu_{i}^{2}\right)  $ where $E,\nu$ are Young's modulus and Poisson's ratio
of each elastic body. It turns out that for $\alpha>0.57$, there is a
spontaneous snap into full contact, and from this state, detachment should
occur only at values of stress close to theoretical strength. We shall return
to this parameter, as we have recently generalized it for a multiscale
surface  --- in what we called the "Generalized Johnson Parameter" (GJP) model
(Ciavarella \&\ Papangelo, 2017a). Turning back to historical developments,
JKR was found to be the correct limit for high Tabor parameters in the case of
a sphere of radius $R$ (Tabor, 1977)
\begin{equation}
\mu=\left(  \frac{Rw^{2}}{E^{\ast2}\Delta r^{3}}\right)  ^{1/3}=\left(
\frac{Rl_{a}^{2}}{\Delta r^{3}}\right)  ^{1/3}\label{tabor}%
\end{equation}
where $l_{a}=w/E^{\ast}$ is an alternative way to measure adhesion as a length
scale, and $\Delta r$ is the range of attraction, which for a crystal is of
the order of atomic spacing $a_{0}$. The DMT approximations originally
developed for the spherical problem (Derjaguin et al., 1975), instead holds
for $\mu\rightarrow0$. In DMT, the contact is assumed to be split into
"repulsive" contact areas and "attractive" contact areas, and no effect of
tensile tractions occurs within the repulsive contact area. This opens the
possibility to apply sophisticated solutions using the DMT idea to Persson's
detailed approximate solution of the repulsive problem (Persson \&\ Scaraggi,
2014), and we shall return on this model to make some comparisons and
validations. DMT leads to possibly serious errors for cylinders and spheres,
unless the Tabor parameter is really low, and it is still quite unclear what
is the degree of approximation in general rough contact (Ciavarella, 2017).

If the simple problem of the sphere has taken so long to be completely
understood, it is not surprising that the case with roughness, starting from
the ideal case of two nominally flat surfaces, is still quite remote from
being solved. The effect of roughness is in fact quite less obvious than it
seems. For very soft and large bodies (JKR\ regime), and special types of
roughness, the effect of roughness could be even to enhance adhesion instead
of reducing it as it is common (Guduru, 2007), but this remains a rather
special condition. However, at the other extreme, like at nanoscales and for
hard solids, very simple equations like Rumpf-Rabinowicz (Rumpf, 1990,
Rabinovich et al., 2000) assuming no elasticity at all, work very well for the
spherical geometry, and show large reduction with very small amplitude of
roughness. These findings are for example confirmed by extensive AFM
experiments by Jacobs et al (2013) where from atomic corrugation up to a few
nm, the measured work of adhesion was found to decrease by more than an order
of magnitude. This naturally also raise the very delicate issue of what is
\textquotedblleft work of adhesion\textquotedblright\ in experimental
measurements of adhesion forces.

This significant dependence on rms amplitude of roughness is relatively in
agreement with classical theories which attempt to consider roughness with
"asperities" and seemed to have been confirmed in simple experiments with low
modulus materials like smooth rubber lenses against roughened surfaces (Fuller
and Tabor, 1975). However, classical asperity models like those of Fuller and
Tabor are questionable for the modern view of "random fractal" surfaces, and
more in general it is a problem that no reliable estimates can be made of
quantities like real contact area, mean slope or mean curvature of surfaces
(see\ Ciavarella and Papangelo, 2017b): the problem of sensitiveness to "small
scale" truncation creates a big effect in asperity models, as their adhesion
stickiness parameter includes the radius of asperities, which clearly would go
to zero in the fractal limit, \textit{making stickiness impossible for any
fractal dimension (in the fractal limit)}, a result which is contradicted also
in the models we are about to discuss.

\bigskip Unfortunately, with the advent of fractal surfaces, the emphasis
shifted from the well defined rms amplitude of asperities heights $h_{rms}$,
which is the most easily measured quantity, and in most common situations
reflects the reduction of stickiness, to much more sophisticated quantities,
and no theory at present is able to clarify the general behaviour of rough
contacts under adhesive conditions, not even in terms of orders of magnitude
estimate, which justifies the attempts the two simple models we are about to
discuss, BAM (Bearing Area Model, Ciavarella, 2017a), and GJP (Ciavarella
\&\ Papangelo, 2017a), which at least permit a full exploration of the
parameters of the problem to be done analytically with simple equations. 

Numerical investigations are also extremely demanding and still not very many
in the literature, with the notable exception of Pastewka \&\ Robbins (2014)
who consider self-affine rough surfaces with 3 orders of magnitude in
wavelengths from nano/atomic scale wavelengths to microscale: however,
Pastewka \&\ Robbins (2014, PR in the following) seemed to reach quite
strikingly strange conclusion: that stickiness (as indicated by the area-load
slope) should be independent on rms amplitude of asperities heights $h_{rms}$,
and only depend on a parameter which combines the rms slopes and rms
curvatures. This conclusion was examined by the present authors in a number of
papers (Ciavarella, 2016, 2017a,b, Ciavarella and Papangelo, 2017c,
Ciavarella, Papangelo and Afferrante, 2017), but perhaps the problem was just
that they looked at loading curves, instead of the much more significant
unloading curves, especially since their data on pull-off in the Supplementary
Material part of the paper do not seem to indicate the same dependence of
"stickiness" on rms quantities. Indeed, we have used mostly the data on
pull-off in subsequent modelling. Although it is true that, as remarked by PR,
their pull-off data decay (Fig. S3) do \textit{not} correlate well with
classical Fuller-Tabor asperity model predictions by various orders of
magnitude, it is also true that they do not correlate well with the PR
parameter (actually worse), and this motivated the implementation of the
BAM\ and GJP models.

The present paper therefore has three purposes:-

1) to review and further validate BAM and GJP with a new set of data, namely
with Persson and Scaraggi (2014) paper

2) to compare BAM and GJP also outside the range of conditions for which they
were both validated.

3) to suggest possible further research by making a number of observations,
for example about force-separation laws

Notice that the GJP (Generalized Johnson Parameter, Ciavarella \&\ Papangelo,
2017a), is not a proper "model", but simply postulates that pull-off should
depend on a single parameter, by analogy with a single sinusoid case as we
have discussed above. The actual dependence on this generalized $\alpha$ was
simply fitted to the Pastewka-Robbins data, and found to be exponential, but
even in these original data, the approximation was not excellent (despite
better than any other single parameter with which we compared). Therefore, we
expect some significant error is possible, whereas the BAM model, being based
on a reasonable DMT theory, should give results which are comparable to
Persson and Scaraggi (2014) paper, at a much reduced cost, and the possible
errors would require in general a very sophisticated numerical investigation,
similar or better than  Pastewka \&\ Robbins (2014, PR), which is way beyong
the scope of the present paper, and in general quite demanding.

\section{A short review of BAM and GJP models}

\subsection{BAM}

In a recent note, the BAM (Bearing Area Model) model was introduced by
Ciavarella (2017a), a single-line equation estimate for adhesion between
elastic (hard) rough solids with Gaussian multiple scales of roughness. BAM
starts from the observation that the entire DMT solution for \textquotedblleft
hard\textquotedblright\ spheres (Tabor parameter tending to zero) assuming the
Maugis law of attraction, is very easily obtained using the Hertzian
load-indentation law and estimating the area of attraction as the increase of
the bearing area geometrical intersection when the indentation is increased by
the Maugis range of attraction (see Fig.1).

\begin{center}%
\begin{tabular}
[c]{ll}%
{\includegraphics[
height=2.1121in,
width=3.1474in
]%
{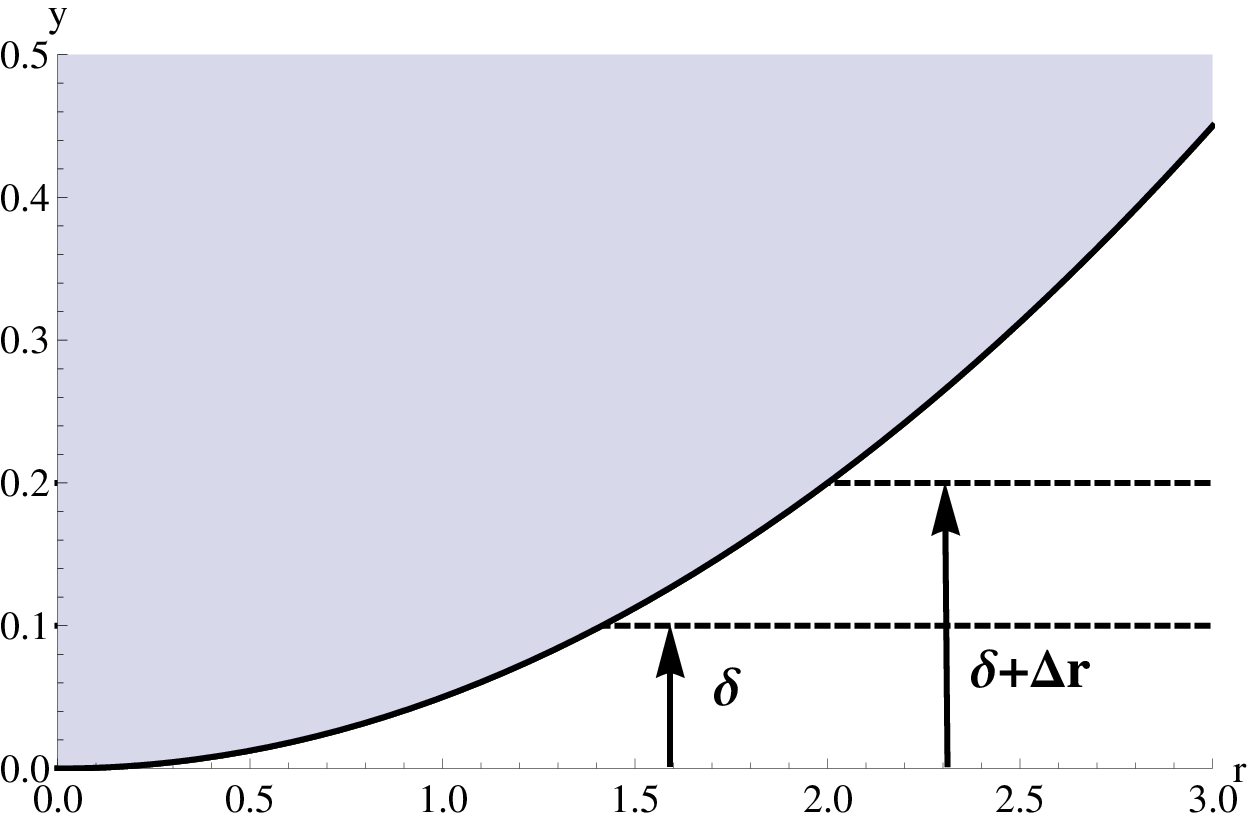}%
}
& (a)\\%
{\includegraphics[
height=2.0158in,
width=3.1474in
]%
{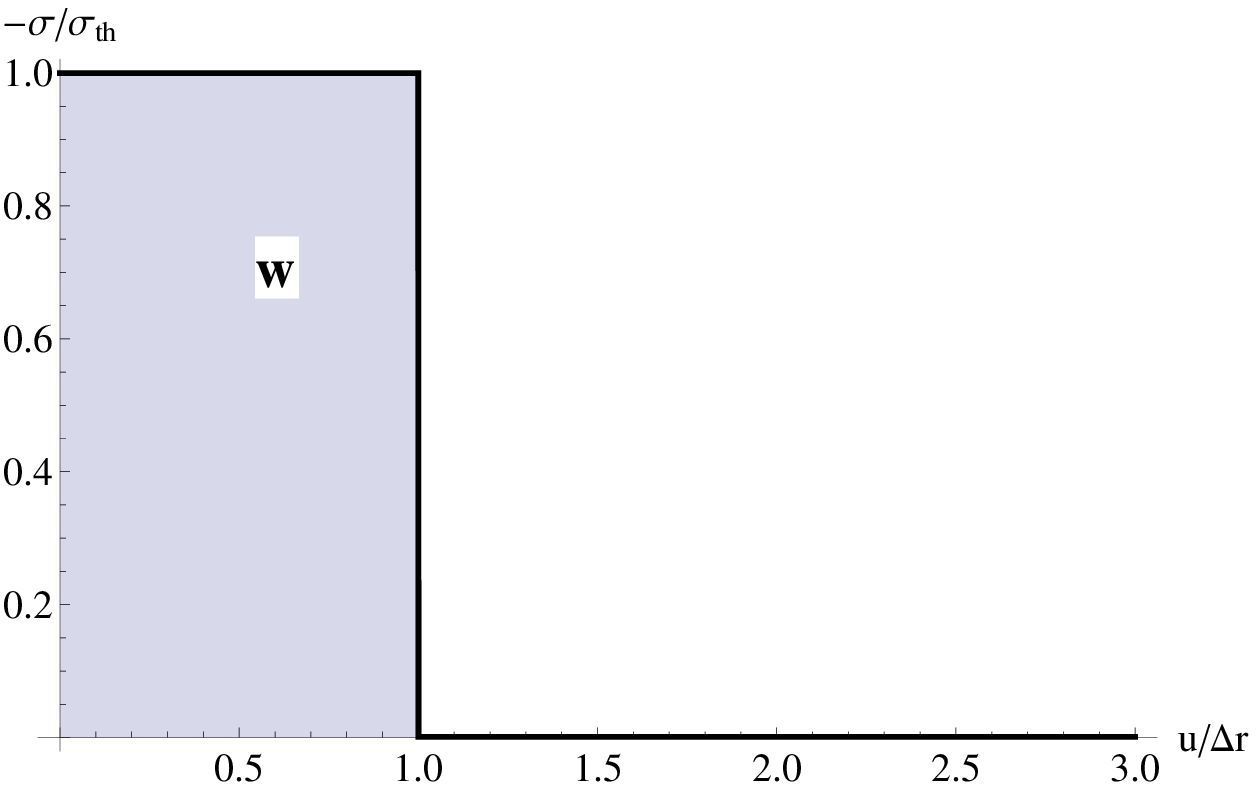}%
}
& (b)
\end{tabular}

Fig.1 - (a) A parabolic elastic body in adhesive contact with a rigid plane;
(b) Maugis forces of attraction \emph{("Maugis-Dugdale potential")}
\end{center}

\bigskip

BAM shows that adhesion, as already well known, for hard solids at macroscopic
scale is destroyed quite easily and the problem remains that contact area is a
ill-defined \textquotedblleft magnification\textquotedblright\ dependent
quantity (Ciavarella and Papangelo, 2017b). When elastic modulus decreases
sufficiently, observable adhesion may be possible, although then the
assumptions of BAM may become questionable, in particular those of the DMT
behaviour, which for a sphere would be low Tabor parameter --- whereas Tabor
parameter for a multiscale roughness can only be estimated, perhaps at the
smallest scale, although this itself is not a solid and unique definition.

\bigskip The BAM  model, summing up repulsive\ (coming from Persson's (2007)
solution, corrected in a prefactor as in Papangelo, Hoffmann and Ciavarella
(2017)) and attractive contributions which is estimated purely from
geometrical considerations, in the case of low fractal dimension (which is
where Persson's solution is simplest and shows no dependence on the small
scale details, but is also the case of most practical interest, see Persson et
al., 2005), gives
\begin{equation}
\frac{\sigma\left(  u\right)  }{\sigma_{th}}\simeq\frac{3}{8\gamma}%
q_{0}h_{rms}\frac{E^{\ast}}{\sigma_{th}}\exp\left(  \frac{-u}{\gamma h_{rms}%
}\right)  -\frac{1}{2k}\times\left[  Erfc\left(  \frac{u-\Delta r}{\sqrt
{2}h_{rms}}\right)  -Erfc\left(  \frac{u}{\sqrt{2}h_{rms}}\right)  \right]
\label{magic-formula}%
\end{equation}
where $\frac{\sigma\left(  u\right)  }{\sigma_{th}}$ is the ratio between the
actual stress between the surfaces (compressive as positive) with respect to
theoretical stress; $q_{0}$ is the short wavevector for a power-law fractal
self-affine profile (a more general definition of the BAM model would require
the use of the entire Power Spectrum Density of a surface, but this is not a problem).

This closed form result for the entire curve of pressure vs mean separation
obviously results in a pull off if we find numerically the minimum as a
function of $u$. The equation depends only on $h_{rms},q_{0}$ and no other
aspect of Power Spectrum, so the pull-off depends only on these two
quantities, and not on small scale details. In particular, notice that using
the constant $k\simeq2$ which is imposed to attempt to fit the complicated
shape of adhesive zones in a rough contact which are rather elongated (see
Pastewka-Robbins, 2014) and also to cover intermediate Tabor parameters, comes
at the expense of modelling the very low $h_{rms}$ as obviously the limit
becomes $\frac{\sigma\left(  u\right)  }{\sigma_{th}}\simeq-\frac{1}{k}$ and
not $-1$. The comparison with Pastewka-Robbins, 2014 data was rather
satisfactory, as we shall see again when we make further comparisons below.

\subsection{\bigskip GJP}

In another note (Ciavarella \&\ Papangelo, 2017a), we introduced a
"generalized Johnson's parameter", which is
\begin{equation}
\alpha\left(  \zeta\right)  =\frac{w}{U_{el}\left(  \zeta\right)  }%
=\frac{l_{a}}{\frac{\pi}{2}\int_{q_{0}}^{q_{1}}q^{2}C\left(  q\right)
dq}=\frac{l_{a}}{l\left(  \zeta\right)  }\label{alfa}%
\end{equation}
where is indeed the generalized Johnson parameter for a multiscale surface,
since the elastic strain energy to flatten the surface, $U_{el}\left(
\zeta\right)  $, depends on the entire Power Spectrum Density (PSD) of the
(isotropic) rough surface $C\left(  q\right)  $, where the surface is
considered up to the magnification\footnote{The important point is that this
"truncation" which in mathematical terms is perfectly fine, but in practise is
quite arbirary assumption typical of many "modern" fractal theories including
those of Persson for which we can look at a problem at different
"magnifications", will not be always needed, as in the most important
practical class of problems, those of low fractal dimension, this parameter
converges.} $\zeta=q_{1}/q_{0}$ Here $q_{0},q_{1}$ are the low cutoff and high
wavevector cut-off of the ideal power-law spectrum (more general spectra
require no difficult generalization). Notice that we have introduced an
effective length of adhesion $l\left(  \zeta\right)  $, and see the derivation
of eqt.8-9 of Persson (2002) for $U_{el}\left(  \zeta\right)  $.

Introducing this parameter, does not solve the problem. The novelty of the GJP
note was the \textit{postulate }that, as in sinusoidal case the pull-off value
depends mainly on $\alpha\ $(at intermediate range of Tabor parameters), the
multiscale problem will depend mostly on a generalized Johnson $\alpha$
parameter. A comparison with other known single parameters (that of Fuller and
Tabor, that of Pastewka and Robbins, and that of the $h_{rms}$ alone), proved
that indeed this postulate was the best. We did not make a full comparison
however with the BAM model. Indeed, for power law PSD and the usual case of
$H>0.5$ (low $D$) the integral converges quickly (see the original paper for
details), and takes the limit value%
\begin{equation}
l\left(  \infty\right)  _{lowD}=\frac{\pi}{2}\frac{h_{0}^{2}}{\lambda_{0}%
}\frac{H}{2H-1}%
\end{equation}
which gives a very gentle dependence on Hurst exponent: the energy is mainly
stored in the long wavelength components. Notice that there is no true
threshold below which surfaces are "absolutely sticky" (probably because of
energy barriers), whereas in practical terms we can define a threshold for
them to be "absolutely unsticky".

\section{\bigskip Comparisons}

For the typical fractal dimension $D=2.2$ ($H=0.8$), GJP becomes
\begin{equation}
\alpha=\frac{3}{4}\frac{2w/E}{q_{0}h_{rms}^{2}}%
\end{equation}
and from the Ciavarella \&\ Papangelo (2017a) fit,
\begin{equation}
\log\left\vert \frac{\sigma_{\min}}{\sigma_{th}}\right\vert =-1.62-\frac
{2.14}{\alpha}=-1.62-\frac{2.14}{0.6}0.8\frac{q_{0}h_{rms}^{2}}{2w/E}%
\end{equation}
which in particular for the majority of PR results has $w/E=l_{a}=0.05a_{0}$,
whereas $q_{0}=2\pi/\left(  2048a_{0}\right)  $
\[
\log\left\vert \frac{\sigma_{\min}}{\sigma_{th}}\right\vert
=-1.62-0.088\left(  \frac{h_{rms}}{a_{0}}\right)  ^{2}%
\]
whereas for $w/E=l_{a}=0.005a_{0}$,
\[
\log\left\vert \frac{\sigma_{\min}}{\sigma_{th}}\right\vert =-1.62-0.88\left(
\frac{h_{rms}}{a_{0}}\right)  ^{2}%
\]
and BAM and GJP are compared in Fig.2: solid lines are GJP predictions, and
cross symbols are for BAM (red for $l_{a}=0.05a_{0}$, blue for $l_{a}%
=0.005a_{0}$), and notice that we have included also data for larger fractal
dimensions (different symbols circles, squares and triangles indicate the
fractal dimension), as well as open symbols which indicate larger rms slope
$h_{rms}^{\prime}=0.3$ (closed symbols are for $h_{rms}^{\prime}=0.1$). The
fit of both BAM and GJP is reasonable, despite there seems to be a certain
deviation for the low adhesion case for $l_{a}=0.005a_{0}$ and the effect of
rms slopes is not entirely clear.

Notice that both GJP and BAM would have a different prediction for larger
fractal dimension, and while GJP has been tested also to include this effect
(and has shown again reasonable prediction, within the limited number and
possible accuracy of data), BAM would require the implementation of the fuller
Persson's solution and this has not been done, also because it is of limited
practical interest.

\begin{center}%
\begin{tabular}
[c]{ll}%
{\includegraphics[
height=3.0369in,
width=5.056in
]%
{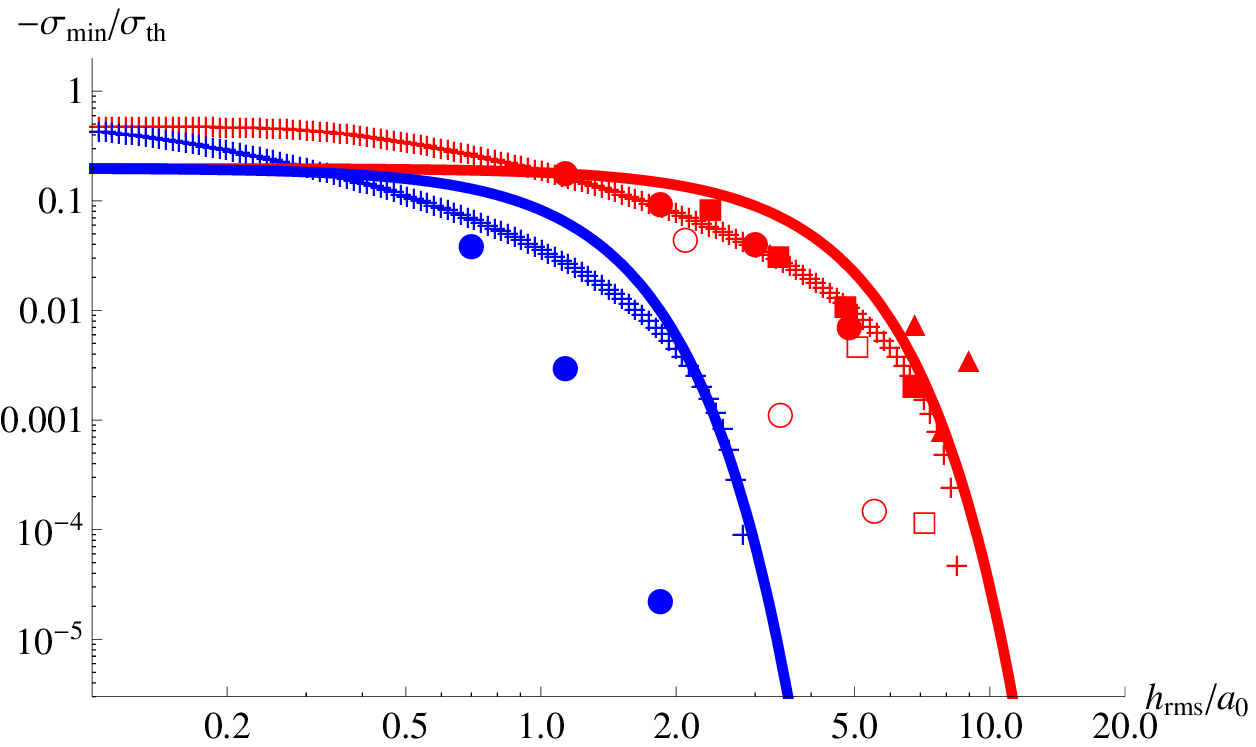}%
}
&
\end{tabular}

Fig.2. Pull-off value decay with parameter. Data are shown in PR\ paper with
the same symbols as they will be shown here, so $h_{rms}^{\prime}=0.1,0.3$
(closed, open symbols), and for $l_{a}/a_{0}=0.05,0.005$ (red, blue) --- we
omit the change in size of the symbols since $q_{s}/a_{0}$ increasing from $4$
to $64$ also corresponds to an increase of $h_{rms}$ which is easy to follow
in the diagram\footnote{Surprisingly, in Fig.S3 there are some blue closed
symbols ($h_{rms}^{\prime}=0.1$ and $l_{a}/a_{0}=0.005$), which appear
curious, as they appear as non-sticky in Fig. 4 of the paper. Also, the
fractal dimension in Fig. S3 does not appear in correct order, as low $H$ seem
to have higher rms amplitude, whereas the opposite trend should occur.
Probably there is an inversion of the data for $H=0.3$ and $H=0.8$, which is
however irrelevant for the present scopes.}.
\end{center}

\subsection{Extrapolations}

We remain with the "Lennard-Jones" estimate of the potential, $w/E=l_{a}%
=0.05a_{0}$, but we now vary the small wavevector $q_{0}=2\pi/\left(
2048xa_{0}\right)  $ where $x$ is a variable to vary the PR case. In the GJP
model, we obtain using the same fit of the PR data, and
\begin{equation}
\log\left\vert \frac{\sigma_{\min}}{\sigma_{th}}\right\vert \simeq
-1.62-\frac{179}{2048}\frac{1}{x}\left(  \frac{h_{rms}}{a_{0}}\right)  ^{2}%
\end{equation}
: $8.\,\allowbreak740\,2\times10^{-2}$and this clearly shows that for a given
rms amplitude of roughness, but increasing largest wavelength (which means
"slope" at macroscopic scale), stickiness increases, as it is intuitive.
Further, given the shape of the curve, we can assume that stickiness is
exhausted when $\frac{\sigma_{\min}}{\sigma_{th}}=10^{-5}$ or $\log
10^{-5}=\allowbreak-11.5$, giving
\[
\left(  \frac{h_{rms}}{a_{0}}\right)  _{thresh}\simeq10.61x^{1/2}%
\]
which can be compared with the estimate made with the BAM model, which instead
requires a numerical routine to find the minimum of the BAM equation.

The comparison is shown in Fig.3, where we see that BAM (solid black curve)
and GJP (solid red line) compare reasonably well in terms of threshold, but
differ considerably at intermediate values. The disagreement at very low rms
amplitudes also comes from having assumed a simple exponential fit and was
judged reasonable given that amplitudes below the lattice spacing $a_{0}$ do
not make any sense.

\begin{center}%
\begin{tabular}
[c]{ll}%
{\includegraphics[
height=3.0369in,
width=5.056in
]%
{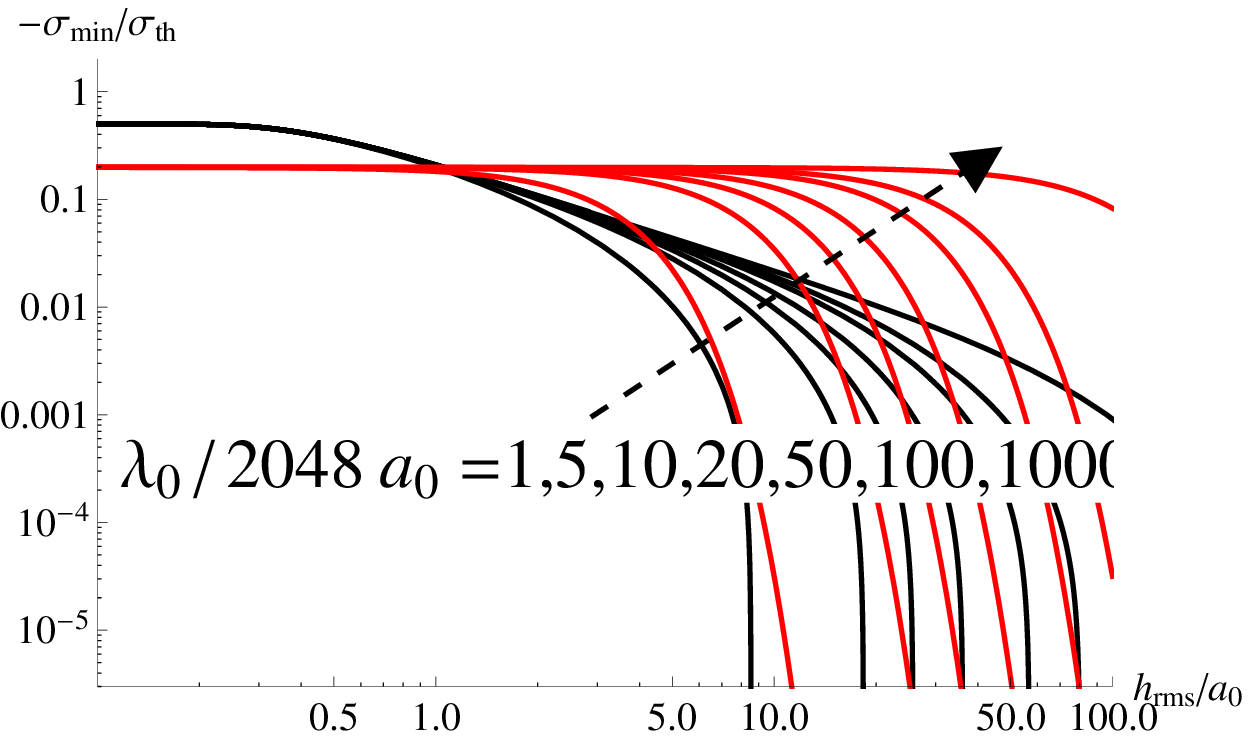}%
}
&
\end{tabular}

Fig.3. Pull-off value decay with different wavelength $\lambda_{0}$ (black
solid lines are BAM model, red solid lines GJP) While the threshold for
stickiness is almost perfectly coincident for the two models, significant
deviations occur for the prediction of actual pull-off in the sticky region
\end{center}

\section{\bigskip Further validations and comparisons}

In the BAM\ model, we have assumed a Maugis type of potential, because that
becomes very convenient for the extension to the random rough surfaces. In the
classical spherical problem, the exact details of the force law are not
important, and this was explained in very nice details in Barber (2013) due to
similarity considerations in quadratic profiles for the rigid body, but it
translates to a very good approximation also in the case of elastic spheres
(but we mantain homogeneous halfspace). However, some caution should be
exercised when modelling non-quadratic profiles. For a perfectly flat surface,
i.e. in the limit of negligible roughness, indeed, the force law gives exactly
the force for a given separation, and the details of the force give the actual
minimum (theoretical strength), for a given surface energy (which is after all
simply a measure of the integral of force-law from zero to infinity).

This should also be borne in mind in future simulations, where the details of
the potential will matter. Pastewka \& Robbins (2014) use a truncated spline
representation of their force-law, which we have attempted to model with a
Maugis law in the BAM model.

Persson \&\ Scaraggi (2014) introduce a pure DMT solution, which makes use of
the elaborate Persson's solution for repulsive (adhesionless) contact, which
includes the full distribution of separations (with multiple recursive
integrals have to be done, and then further convoluted with the force
separation law, i.e. integrated, to estimate the attractive force), and not
just the macroscopic force-separation law as we have used in BAM. They use a
purely adhesive potential (since in DMT the repulsion is taken care by
Signorini boundary condition with zero separation in the contact zone) which
imitates the Lennard-Jones potential%
\begin{equation}
p_{a}^{+}\left(  u\right)  =B^{+}\left[  \left(  \frac{\Delta r}{u+\Delta
r}\right)  ^{n}-\beta\left(  \frac{\Delta r}{u+\Delta r}\right)  ^{m}\right]
\end{equation}
and most of their results are for $\beta=0$, where $B^{+}$ is adjusted to make
the integral equal to the given surface energy. Here, $n=3$ resembles van der
Waals attractive law. Obviously for $\beta=0$, $p_{a}^{+}\left(  0\right)
=B^{+}$ is by construction the theoretical strength, and since $B^{+}=\frac
{w}{\Delta r}\left(  n-1\right)  $ there is a very significant (linear)
dependence on the value of $n$. Notice that for Maugis as we use in BAM,
$\sigma_{th}\simeq\frac{w}{\Delta r}$ and this would seem to correspond to
$B^{+}$ when $n=2$ --- however, this is obviously only the correspondence of
the peak values, but when roughness will be present, details may matter. But
we shall see that this coincidence seems indeed to correspond to much better
agreement of the full Persson-Scaraggi DMT model with the simple BAM.

For their calculations, Persson \&\ Scaraggi use $\Delta r=4\times10^{-10}m$,
and $E^{\ast}=\frac{4}{3}TPa$, so notice that $\frac{l_{a}}{\Delta
r}=w/\left(  E^{\ast}\Delta r\right)  =0.1/\left(  \frac{4}{3}10^{12}%
\times4\times10^{-10}\right)  =\allowbreak1.88\times10^{-4}\ $which is 2
orders of magnitude smaller the value expected for Lennard Jones of 0.05. This
law in the case of $\beta=0$ introduces a spike at "zero separation"
(theoretical strength $\sigma_{th}$) which would tend to increase the adhesive
force, given in a state of repulsive contact there are significant regions at
near zero separations and indeed the probability distribution \textit{is
singular there} (see Figure 9 of Almqvist et al., 2011). Their comparison with
Persson's own JKR theory seems to indicate a very significant difference (and
the JKR theory seems limited to positive loads anyway, so there is no
comparison for the most important part of an adhesive solution) -- part of
this difference is intrinsic in the assumptions: in particular, Fig. 6 shows
contact area results, where for DMT the area is purely repulsive, and for JKR
it is both repulsive and attractive. However, in some cases this is not
sufficient to explain why repulsive DMT areas are sometimes higher than the
total JKR areas.\ A\ full check of DMT solution is missing in their paper
because the full numerical solution is limited to very small bandwidths,
presumably as computational cost for a true multiscale surface solved with a
non-linear adhesive BEM code is still too demanding for present computers, and
indeed Pastewka \&\ Robbins (2014), are probably the most advanced still
today.\ Unfortunately, Persson and Scaraggi's method is not trivial to
implement as is BAM, nor the code is available to the public. However, we can
make some qualitative comparisons using the calculations they report in a few
cases. Their surfaces have roll-off and this would require using more
sophisticated version of Persson's repulsive solution. Indeed, the most
important contribution to roll-off is to make surfaces more Gaussian, and it
does not contribute significantly to the rms amplitude nor to the repulsive
solution, so we can neglect the roll-off region and use the pure power-law
solution, for the sake of simplicity. Hence, low wavevector is $q_{0}=10^{6},$
rms amplitude is $h_{rms}=6\times10^{-10}m$. With the GJP model, we obtain
\begin{equation}
\log\left\vert \frac{\sigma_{\min}}{\sigma_{th}}\right\vert =-1.62-\frac
{2.14}{0.6}0.8\frac{10^{6}\left(  0.6\times10^{-9}\right)  ^{2}}{2}\frac{4}%
{3}10^{12}\frac{1}{w}=-1.62-0.684/w
\end{equation}
whereas BAM requires to find numerically the minimum of the force-separation.
A comparison of the predictions with the Persson and Scaraggi's results is in
Tab.1, where we see that GJP prediction is actually rather poor (there are 1-2
orders of magnitude differences with higher pull-off in the Persson-Scaraggi's
estimate) whereas BAM is rather close and very reasonable, considering the
much simpler implementation of a single closed form equation. Notice we have
removed the factor $k=2$ of the original BAM model, which results in a more
realistic trend towards the very low amplitudes of roughness, tending to
theoretical strength. However, notice that the fit of Persson-Scaraggi's data
would improve for some data, and not for others.

\begin{center}%
\begin{tabular}
[c]{ll}%
{\includegraphics[
height=3.0369in,
width=5.056in
]%
{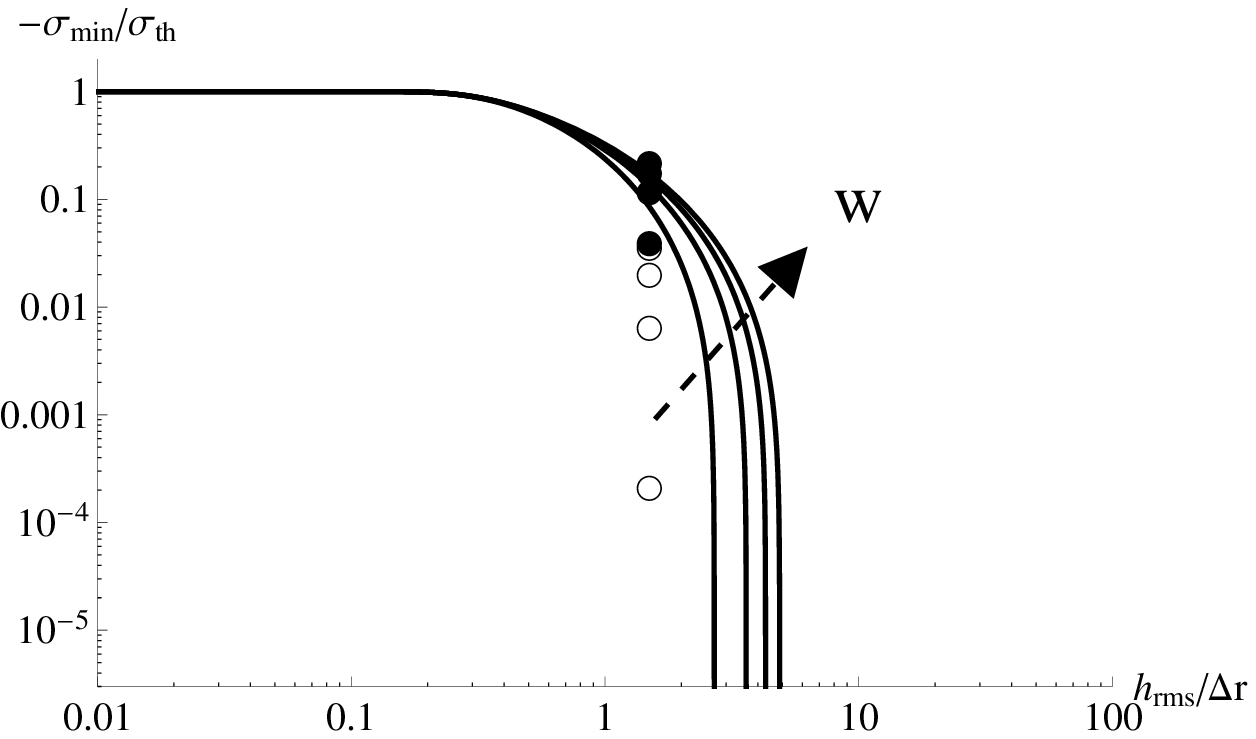}%
}
&
\end{tabular}

Fig.4. Pull-off value decay for results from Persson \&\ Scaraggi (2014) Fig.7
for different surface energies $w$ (solid filled circles), and comparison with
BAM (black solid lines) with reasonable agreement. Instead, comparison with
GJP (empty circles) shows less good agreement.%

\begin{tabular}
[c]{||l||l||l||l||l||}\hline\hline
$w$ [J/m$^{2}]$ & $\sigma_{th}\left[  GPa\right]  $ & $\left\vert \sigma
_{\min}/\sigma_{th}\right\vert \left(  PS\right)  $ & $\left\vert \sigma
_{\min}/\sigma_{th}\right\vert \left(  GJP\right)  $ & $\left\vert
\sigma_{\min}/\sigma_{th}\right\vert \left(  BAM\right)  $\\\hline\hline
0.1 & 0.5 & $\allowbreak0.04$ & $2.12\times10^{-4}$ & $0.08$\\\hline\hline
0.2 & 1 & $0.12$ & $6.5\times10^{-3}$ & $0.132$\\\hline\hline
0.3 & 1.5 & $\allowbreak0.18$ & $2.02\times10^{-2}$ & $0.157$\\\hline\hline
0.4 & 2 & $\allowbreak0.225\,$ & $3.6\times10^{-2}$ & $0.173$\\\hline\hline
\end{tabular}

Tab.1 - Some results from Persson \&\ Scaraggi (2014) Fig.7, and comparison
with GJP and BAM%

\begin{tabular}
[c]{||l||l||l||}\hline\hline
$n$ & $\sigma_{th}\left[  GPa\right]  $ & $\left\vert \sigma_{\min}%
/\sigma_{th}\right\vert \left(  PS\right)  $\\\hline\hline
1.5 & 0.375 & $\allowbreak0.13$\\\hline\hline
2 & 0.75 & $0.16$\\\hline\hline
3 & 1.5 & $\allowbreak0.17$\\\hline\hline
4 & 2.25 & $\allowbreak0.19\,$\\\hline\hline
\end{tabular}

Tab.2 - Some results from Persson \&\ Scaraggi (2014) Fig.8, $w=0.3J/m^{2}$.
For comparison with GJP and BAM see Tab.1 ($\left\vert \sigma_{\min}%
/\sigma_{th}\right\vert \left(  GJP\right)  =2.02\times10^{-2}$ and
$\left\vert \sigma_{\min}/\sigma_{th}\right\vert \left(  BAM\right)  =0.157$)

\end{center}

Tab.2 shows that the choice of the power law in the adhesive force-separation
law is not indifferent in the results obtained by Persson and Scaraggi,
although its effect is not dramatic and the prediction of BAM extremely close
to the case $n=2$ , which incidentally we suspected was the closest to a
Maugis potential in the beginning of the paragraph.

\section{Conclusion}

We have compared two simple models for pull-off of hard elastic solids for low
Tabor parameters, one based on a new geometrical variant of the DMT solution
for the sphere, and another based on the postulate that pull-off should depend
only on the ratio of surface energy and elastic strain energy to flatten the
surface. We have seen that in the range of data from Pastewka-Robbins for
which the data were calibrated, there is obviously very similar predictive
capability. However, outside this range, the difference may be larger, despite
the threshold of stickiness seems to be given by reasonably very close results.

Indeed, validating the models for an independent set of data, those obtained
with the Persson and Scaraggi's DMT\ model, we found that the BAM model seems
very similar to the much more complex Persson and Scaraggi's DMT\ model, and
therefore is most promising. GJP instead, which was an empirical fit after
all, shows worryingly large differences in the case of Persson and Scaraggi's
DMT\ paper, with respect to both BAM and Persson and Scaraggi.

\section{\bigskip\bigskip References}

Almqvist, A., Campana, C., Prodanov, N., \& Persson, B. N. J. (2011).
Interfacial separation between elastic solids with randomly rough surfaces:
comparison between theory and numerical techniques. Journal of the Mechanics
and Physics of Solids, 59(11), 2355-2369.

Barber, J. R. (2013). Similarity considerations in adhesive contact problems.
Tribology International, 67, 51-53.

Bradley R S (1931) The Molecular Theory of Surface Energy. Phil. Mag. 11 p846 -849

\bigskip Ciavarella, M. (2017). On the use of DMT approximations in adhesive
contacts, with remarks on random rough contacts. Tribology International, 114, 445-449.

Ciavarella, M. (2016). On a recent stickiness criterion using a very simple
generalization of DMT theory of adhesion. Journal of adhesion science and
Technology, 30(24), 2725-2735.

Ciavarella, M. (2017a) A very simple estimate of adhesion of hard solids with
rough surfaces based on a bearing area model. Meccanica, 1-10. DOI 10.1007/s11012-017-0701-6

Ciavarella, M. (2017b). On Pastewka and Robbins' Criterion for Macroscopic
Adhesion of Rough Surfaces. Journal of Tribology, 139(3), 031404.

Ciavarella, M., \& Papangelo, A. (2017a). A generalized Johnson parameter for
pull-off decay in the adhesion of rough surfaces, Physical Mesomechanics
\textperiodcentered\ December 2017

Ciavarella, M., \& Papangelo, A. (2017b). Discussion of \textquotedblleft
Measuring and Understanding Contact Area at the Nanoscale: A
Review\textquotedblright\ by Tevis DB Jacobs and Ashlie Martini. Applied
Mechanics Reviews. http://appliedmechanicsreviews.asmedigitalcollection.asme.org/article.aspx?articleid=2658189

\bigskip Ciavarella, M., \& Papangelo, A. (2017c). A modified form of
Pastewka--Robbins criterion for adhesion. The Journal of Adhesion, 1-11.

Ciavarella, M., Papangelo, A., \& Afferrante, L. (2017). Adhesion between
self-affine rough surfaces: Possible large effects in small deviations from
the nominally Gaussian case. Tribology International, 109, 435-440.

Ciavarella, M., Greenwood, J. A., \& Barber, J. R. (2017). Effect of Tabor
parameter on hysteresis losses during adhesive contact. Journal of the
Mechanics and Physics of Solids, 98, 236-244.

Derjaguin, B. V., Muller V. M. \& Toporov Y. P. (1975). Effect of contact
deformations on the adhesion of particles. J. Colloid Interface Sci., 53, pp. 314--325

Fuller, K. N. G., \& Tabor, D. The effect of surface roughness on the adhesion
of elastic solids. Proc Roy Soc London A: 1975; 345:1642, 327-342

Guduru, P.R. (2007). Detachment of a rigid solid from an elastic wavy surface:
theory J. Mech. Phys. Solids, 55, 473--488

Jacobs, T. D., Ryan, K. E., Keating, P. L., Grierson, D. S., Lefever, J. A.,
Turner, K. T., ... \& Carpick, R. W. (2013). The effect of atomic-scale
roughness on the adhesion of nanoscale asperities: a combined simulation and
experimental investigation. Tribology Letters, 50(1), 81-93.

Johnson, K. L., K. Kendall, and A. D. Roberts. (1971). Surface energy and the
contact of elastic solids. Proc Royal Soc London A: 324. 1558.

Johnson, K. L. (1995). The adhesion of two elastic bodies with slightly wavy
surfaces. Int. J. Solids Structures, 32\textbf{\ (}No. 3/4\textbf{)}, 423-430.

Johnson KL (1995) The adhesion of two elastic bodies with slightly wavy
surfaces. Int J Solids Struct 32(3/4):423--430

Papangelo, A., Hoffmann, N., \& Ciavarella, M. (2017). Load-separation curves
for the contact of self-affine rough surfaces. Scientific reports, 7(1), 6900.

Pastewka, L., \& Robbins, M. O. (2014). Contact between rough surfaces and a
criterion for macroscopic adhesion. Proceedings of the National Academy of
Sciences, 111(9), 3298-3303.

Persson, B.N.J., Albohr, O., Tartaglino, U., Volokitin, A.I., Tosatti, E.,
(2005). On the nature of surface roughness with application to contact
mechanics, sealing, rubber friction and adhesion. J. Phys.: Condens. Matter.
17, 1--62.

Persson, B. N. J. "Adhesion between an elastic body and a randomly rough hard
surface." (2002) The European Physical Journal E: Soft Matter and Biological
Physics 8, no. 4 : 385-401.

Persson, B. N. J. (2007). Relation between interfacial separation and load: a
general theory of contact mechanics. Physical review letters, 99(12), 125502.

Persson, B. N., \& Scaraggi, M. (2014). Theory of adhesion: Role of surface
roughness. The Journal of chemical physics, 141(12), 124701.

Pastewka L, Robbins MO (2014) Contact between rough surfaces and a criterion
for macroscopic adhesion. Proc Nat Acad Sci 111(9):3298--3303

Rabinovich YI, Adler JJ, Ata A, et al. (2000) Adhesion between nanoscale rough
surfaces: I. Role of asperity geometry. J Colloid Interface Sci 232(1): 10--16.

Rumpf H. Particle Technology. London/New York: Chapman and Hall, 1990.

Tabor, D. (1977). Surface forces and surface interactions. Journal of colloid
and interface science, 58(1), 2-13.

\section{}
\end{document}